\documentclass{webofc}
\usepackage[varg]{txfonts}   % Web of Conferences font
\begin{document}
\title{ALMA Band 9 upgrade: a feasibility study}

\author{\lastname{S. Realini}\inst{1}\fnsep\thanks{\email{realini@astro.rug.nl}} \and
        \lastname{R. Hesper}\inst{1} \and
        \lastname{J. Barkhof}\inst{1} \and
        \lastname{A. Baryshev}\inst{1}
}

\institute{Kapteyn Astronomical Insitute, University of Groningen, Groningen, The Netherlands
            %\and
          }

\abstract{%
We present the results of a study on the feasibility of upgrading the existing ALMA Band 9 receivers (602-720 GHz). In the current configuration, each receiver is a dual channel heterodyne system capable of detecting orthogonally polarized signals through the use of a wire grid and a compact arrangement of mirrors. The main goals of the study are the upgrade of the mixer architecture from Double-Sideband (DSB) to Sideband-separating (2SB), the extension of the IF and RF bandwidth, and the analysis of the possibilities of improving the polarimetric performance.
We demonstrate the performance of 2SB mixers both in the lab and on-sky with the SEPIA660 receiver at APEX, which shows image rejection ratios exceeding 20 dB and can perform successful observations of several spectral lines close to the band edges. The same architecture in ALMA Band 9 would lead to an increase in the effective spectral sensitivity and a gain of a factor two in observation time.
We set up also an electromagnetic model of the optics to simulate the polarization performance of the receivers, which is currently limited by the cross-polar level and the beam squint, i.e. pointing mismatch between the two polarizations. We present the results of the simulations compared to the measurements and we conclude that the use of a polarizing grid is the main responsible of the limitations.
}
\maketitle
\section{Introduction}
\label{sec:intro}
With its 66 high-precision antennas located on the Chajnantor plateau at 5000 m altitude, the Atacama Large Millimeter/submillimeter Array (ALMA) is the largest ground-based aperture synthesis telescope for observation in the millimeter and submillimeter regime. The broad frequency range available (30--950 GHz) is divided into ten different receiver bands, where each receiver unit (called “cartridge”) is built independently. Each receiver band detects two orthogonal linear polarizations and down-converts the signals to an intermediate frequency.
Each of the ten bands should be optimized to provide a good coupling with ALMA telescope as well as low sidelobe and cross-polarization levels. In addition, the beams for the two orthogonal polarizations on the sky should coincide and any deviation, called beam squint, should be minimized.
%The performance related goals are a beam squint $<1$\% of FWHM and a cross-polar level -23 dB below the co-polar total energy level. These requirements are very challenging and not completely fulfilled by all receiver bands, for example Band~9, which is the main topic of the work presented in this proceeding. Due to the non-optimal performance of its receivers, there is an on-going study to make an upgrade to 2SB receivers with a possible improvement of the optical system within the cartridge as described in Section \ref{sec:project}.

Band~9 (610-720 GHz) is ideal to study warm ($T>100$~K) and dense gas spectroscopically for several astrophysical objects: high redshift galaxies, starbursts, black holes in AGN, star-forming regions, proto-planetary disks, and the Solar system. It also allows one to detect high excitation lines of CO, HCN, HNC and HCO+ across cosmic time. Since such observations provide a unique window on galaxy evolution, we should seek the best available detector technique for this band. The present ALMA Band~9 receivers use a double side band (DSB) mixer scheme \cite{Baryshev2015}, while all frequency bands below 650 GHz use dual sideband (2SB) heterodyne receivers to better to remove the atmospheric noise in the image IF band.
Moreover, the existing Band~9 receivers have a cross–polar performance which does not meet the ALMA specifications of -23 dB and they show a relatively large beam squint, which makes this channel not suitable for extended-source polarimetry. 

These issues lead to the definition of a project to study a possible improvement of Band 9 performance within the framework of ESO “Advanced Study for Upgrades of the Atacama Large Millimeter/sub-millimeter Array (ALMA)”. The main goals of the project are described in Section~\ref{sec:project}.

\section{Project overview}
\label{sec:project}

In its current configuration, the ALMA Band~9 receiver is a dual channel heterodyne system capable of detecting orthogonally polarized signals using a wire grid combined with a compact arrangement of mirrors. 
The requirements on the optics performance are a beam squint < 1\% of FWHM and a cross-polar level -23 dB below the co-polar total energy level. However, the cross–polar performance of existing receivers does not meet the requirements specified for the ALMA channels, and they show a relatively large beam squint, which makes this channel not suitable for extended-source polarimetry.

These requirements are very challenging and not completely fulfilled in Band 9. Due to the non-optimal performance of its receivers, we performed a feasibility study to make an upgrade to 2SB receivers with a possible improvement of the optical system within the cartridge. The study, titled “Full 2SB Receiver Upgrade for ALMA Band~9: Implementation Study”, is founded by ESO and the main goals are \cite{report}:
\begin{itemize}
    \item extension of the IF bandwidth to $4\times12$ GHz (2 sidebands and 2 polarizations), without compromising the other performance parameters;
    \item extension of the RF bandwidth beyond the nominal 600-720 GHz;
    \item analysis of the possibility to improve the optical cross-polarization performance compared to the currently installed configuration, e.g. by using a single-horn configuration with an orthomode transducer;
    \item investigation of the causes of the scatter in the beam squint to understand if it could be improved by reducing tolerances of the grid.
\end{itemize}

The technical feasibility of an ALMA Band~9 upgrade from the existing double-sideband configuration to the sideband-separating one has already been demonstrated in a previous ESO study \cite{2sbUpgrade}. Since the proposed rebuilding of the Band~9 cartridges could be an opportunity to bring the cross-polar performance in line with the original specification, we carried out a dedicated analysis to investigate the improvement in the cross-polarization performance by changing the way in which polarization separation is performed, i.e. removing the grid and using an orthomode transducer (OMT) to separate the two orthogonal polarizations.
Here we present also the results of the analysis performed to check the effect of the mounting of the grid on the beam squint.

\subsection{Science drivers}

The proposed upgrade would give significant scientific advantages by improving the sensitivity by a factor 2 through the elimination of the atmospheric noise in the unwanted image sideband. It would also allow a reduction in line confusion in spectral line surveys and a factor 4 improvement in the required integration time. This is highly beneficial because Band~9 requires favorable weather conditions, which only occur a modest fraction of the time.

We identified also additional science cases, especially related to the polarization capability. Among these we find the study of magnetic fields in very dense environments of circumstellar envelopes around evolved stars and high mass star-forming regions through the vibrationally excited water masers at 658 GHz. Polarization observations in Band~9 can be a powerful tool to study also dust settling and grain growth in accretion/protoplanetary disks around young stellar objects.

\section{The ALMA Band 9 receiver}
\label{sec:receiver}
\begin{figure}[h]
% Use the relevant command for your figure-insertion program
% to insert the figure file.
\centering
\includegraphics[scale=0.23]{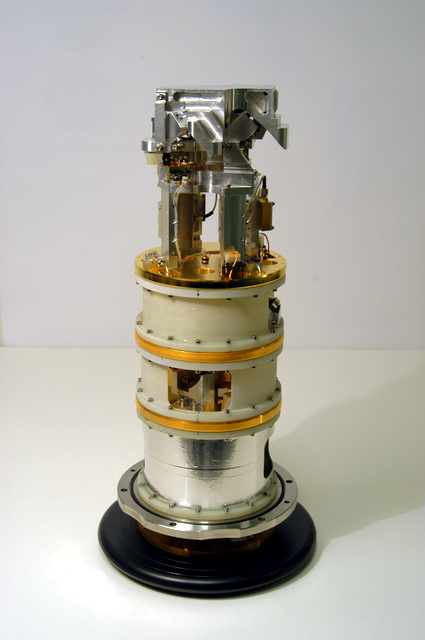}
\hspace{1.2cm}
\includegraphics[scale=0.25]{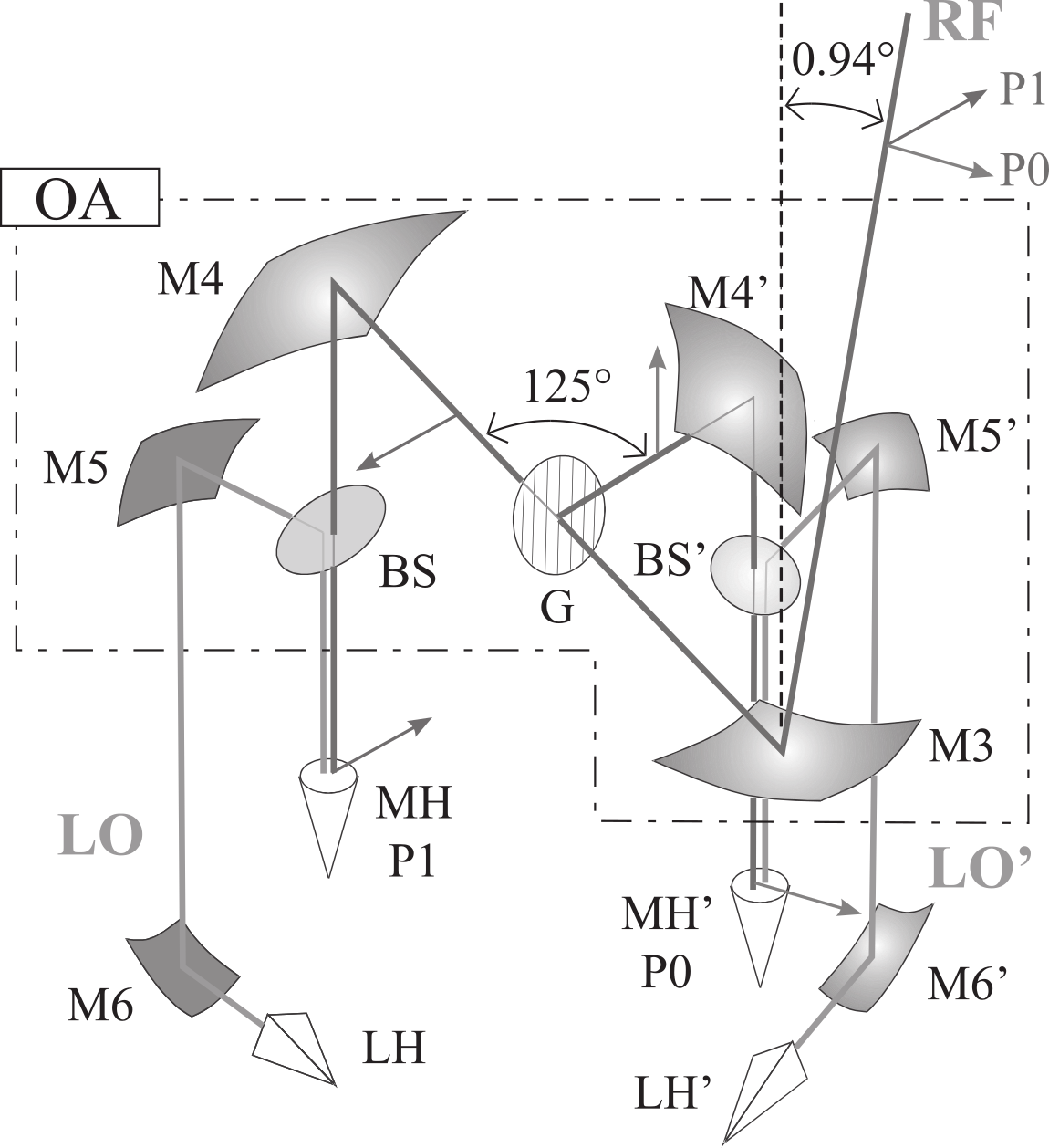}
\caption{\emph{Left:} Picture of ALMA band 9 receiver cartridge. \emph{Right:} Layout of the optical system for focusing the RF beam and injecting the LO signals}
\label{fig:cartridge}       % Give a unique label
\end{figure}

Each ALMA Band 9 receiver unit (Fig.~\ref{fig:cartridge}, left) is installed in the telescope front end and has been designed to detect and down-convert two orthogonal linear polarization components of the light collected by the ALMA antennas \cite{Baryshev2015}. The radiation, collected by the telescope, enters the front end through a vacuum window and a set of infrared filters. The beam is then refocused with a compact arrangement of elliptical mirrors, which is fully contained within the cartridge (Fig.~\ref{fig:cartridge}, right). The optics assembly contains a polarizing grid to separate the two orthogonal linear polarizations and two beam splitters to combine each resulting beam with a local oscillator signal. The two beams are then sent to independent superconductor-insulator-superconductor (SIS) mixers that perform the heterodyne down-conversion. Finally, the generated intermediate frequency (IF) signals are amplified by cryogenic and room-temperature HEMT amplifiers and exported to the telescope’s IF backend for further processing and correlation.

\section{Results}
\label{sec:results}

\subsection{Extension of the IF Bandwidth}
\label{sec:IF_bw}

To determine the maximum IF bandwidth achievable by the Band 9 sideband separating receivers, we should consider the bandwidth of the mixer devices themselves and the technological extensions needed to harness this bandwidth. In addition, even if the mixers can deliver the expected bandwidth, the greatest technical challenge is obtaining cryogenic low-noise amplifiers (LNAs) and quadrature hybrids. Currently, the Observatorio Astronomico de Yebes (Spain) is involved in the further development of this type of amplifiers, especially towards larger bandwidth and better input matching, supported by a dedicated ESO technology development program.

\begin{figure}[h]
\centering
\includegraphics[scale=0.55]{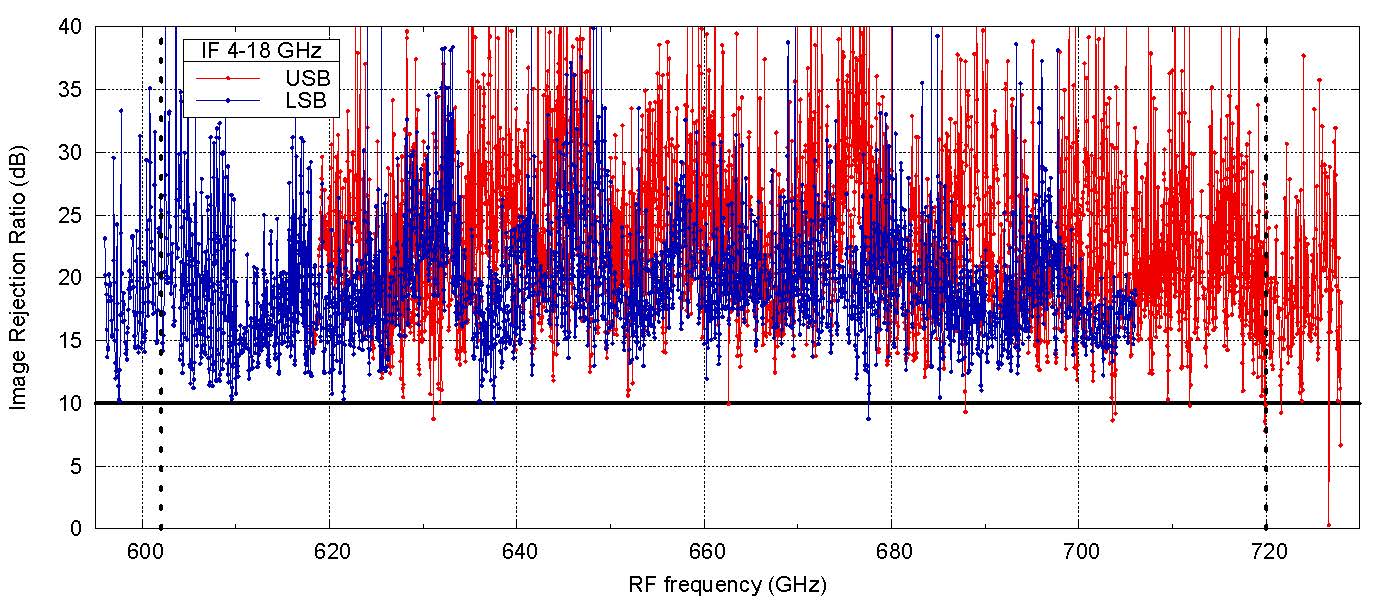}
\caption{Combined IRR data for the tested mixer pair in the IF range 4–18 GHz.}
\label{fig:IRR-IF}
\end{figure}

To test the performance of the sideband-separating mixers, we measured the two most significant properties: the image rejection ratio (IRR) and the noise temperature over the IF band. Figure \ref{fig:IRR-IF} shows the combined IRR data for the tested mixer pair in the IF range 4–18\,GHz.

The measurements show that the existing Band 9 SIS mixer design can achieve a larger IF bandwidth than the original design specification and the performance will be similar up to 18 GHz.

\subsection{Extension of the RF Bandwidth}
\label{sec:RF_bw}
The band edges of ALMA Band 9 were originally chosen to be 602 and 720 GHz because of the presence of strong water vapour absorption lines at 557 and 750 GHz, each pressure broadened of about 50 GHz. However, very dry conditions allow useful observations even outside of this range, especially if the source is sufficiently bright. A good example is a spectral line survey of the Orion-KL star-forming region performed with the SEPIA660 2SB receiver at APEX \cite{sepia-meas}.

As demonstrated by the measurements performed on the SEPIA660 receiver, the existing SIS junctions that make up most of mixer in the delivered Band 9 cartridges typically have an RF bandwidth extending significantly beyond the 602–720 GHz range of the Band 9 specification. Also the reflective optics in the cartridge as well as the corrugated feedhorns have bands wider than the specification. The only modification needed is the extension of the local oscillator (LO) tuning range by the use of commercial components. Therefore, we can extend the RF bandwidth to 580–735GHz.

\subsection{Polarization Performance}
\label{sec:pol}

To evaluate the impact of the polarization grid on the cross-polar level, we simulated the original Band 9 optics using the software GRASP. We built the model according to the values reported in \cite{Granada}. The final layout of the model includes two feedhorns, three ellipsoidal mirrors and the polarization grid. The feedhorn is modelled as a hybrid-mode feedhorn based on the values from the drawings. This is a good approximation to the actual corrugated horn. The mirrors shape is defined by the mechanical drawings assuming the values for the cold structure. The grid is modelled as a regular grid of conducting parallel wires.
We define polarisation 0P as the reflected component from the grid wires, which are parallel to the direction of the horn
axes, while polarisation 1P is the component transmitted through the wire grid.
% A common mirror redirects both polarisation beams towards the sub-reflector mirror with an off-set angle
% of 0.974◦. The two polarization branches are identical, but the reflected branch is rotated 125◦ around an axis
% through the centre of the grid.

\subsubsection{Cross-polarization}
\label{sec:x-pol}
\begin{figure}[h]
% Use the relevant command for your figure-insertion program
% to insert the figure file.
\centering
\includegraphics[scale=0.41]{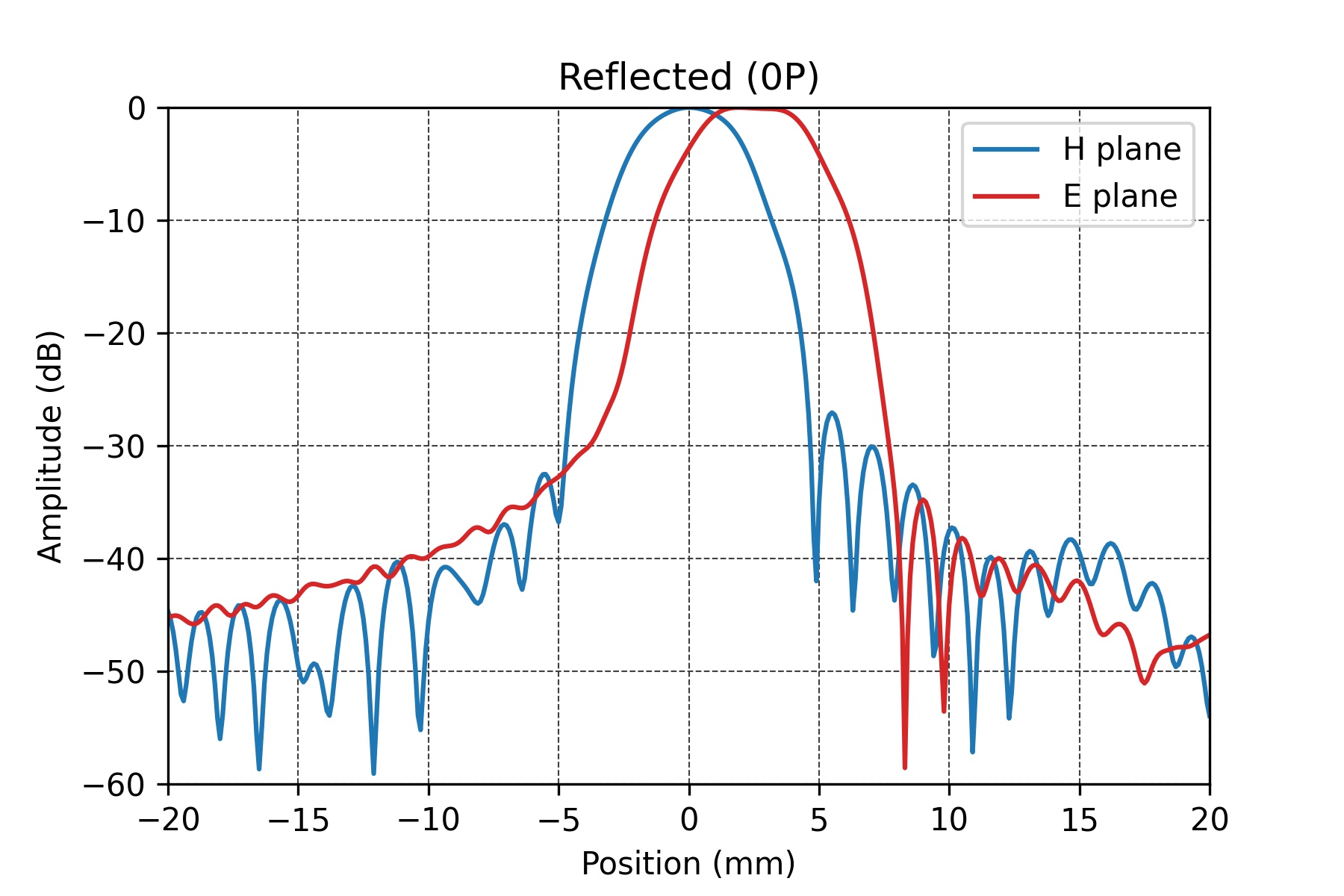}
\includegraphics[scale=0.41]{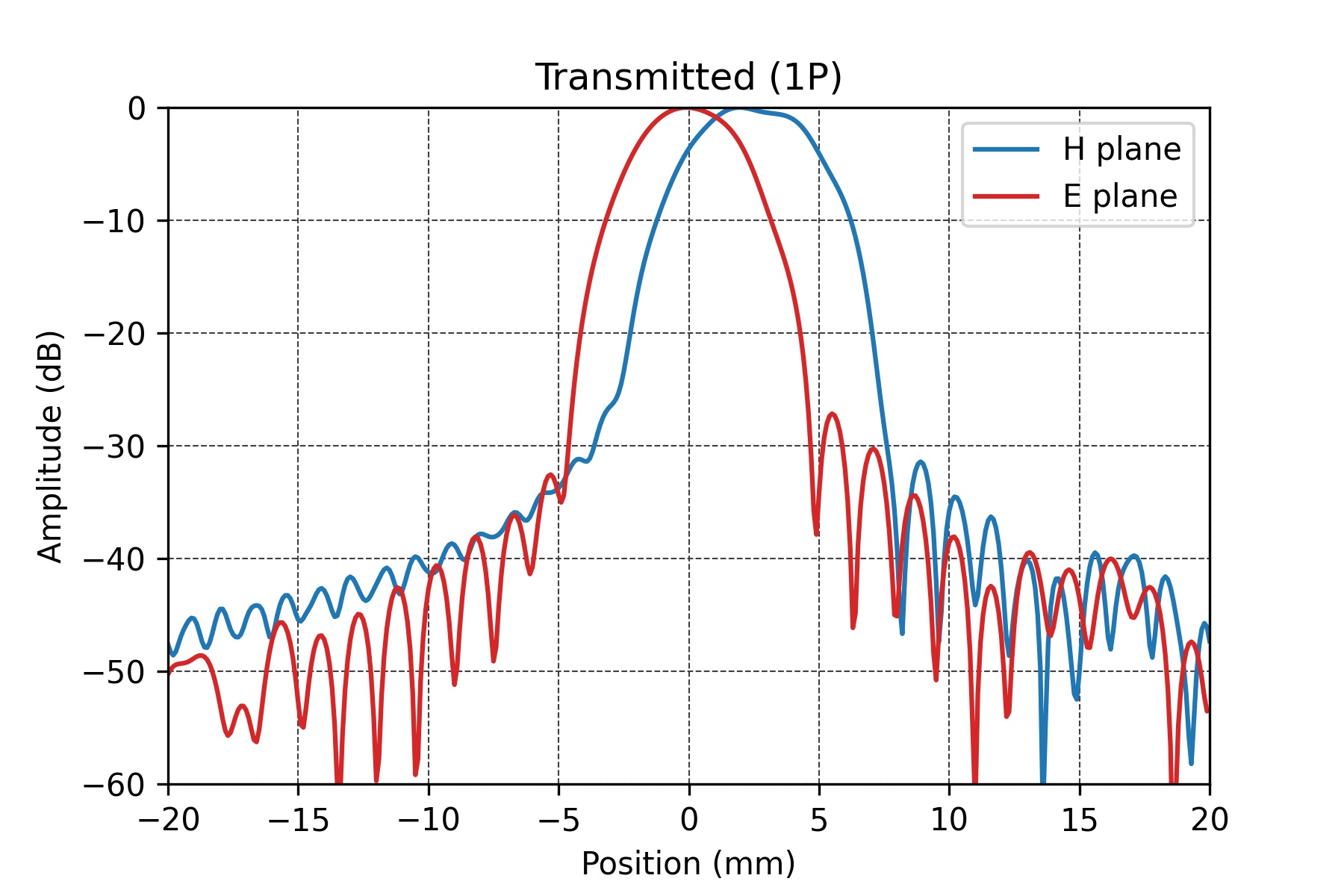}
\caption{Orthogonal cuts of the 0P and 1P polarisation beams in the focal plane. The displacement of the E- and
H-planes depends on the angle of the incoming RF signal, which is 0.94$^\circ$ with respect to the cartridge axis.}
\label{fig:cut}       % Give a unique label
\end{figure}

We started the analysis considering the nominal optics to determine the reference cross-polarization level. We performed main beam simulations using Physical Optics (PO) and Physical Theory of Diffraction (PTD) on all the reflectors. To verify the model, we compared the results of the simulations with the available measured values and we found compatible values of -17 dB. Figure \ref{fig:cut} shows the two orthogonal cuts of the 0P (reflected) and 1P (transmitted) polarisation beams at the same measurement location, i.e. in the focal plane. 
Then we repeated the simulations removing the grid from the model to get an idea of the best cross-pol obtainable with the existing optics when this is re-used in a single-horn configuration. With this configuration we get a cross-polarization level of $\sim$-23.8 dB, i.e. within the specification established for ALMA.

\subsubsection{Beam squint}
\label{sec:squint}

Since the Band 9 orthogonally polarised beams are divided using a wire grid, they follow separate paths from the grid to the related feedhorn. Displacement of the optical components can cause a beam squint between the two main beams on the sky, potentially affecting the receivers calibration procedure. Variations of the beam squint between cartridges are very difficult to calibrate out and must be tightly controlled if not to hamper wide-field polarisation observations.

The beam squint for OMT-based bands is generally within 2\%, while the current configuration of Band 9 with the wire grid has a larger scatter \cite{Neil}. %Figure 5 shows the linear polarisation beam squint (cross-elevation and elevation offset between the Y and X polarisation beams on the sky) in units of percentage of the beam FWHM at the observing frequency for Band 5 and Band 9. 
In addition, the on-sky beam squint shows a preferred direction, with an angle of $\sim$39.5$^\circ$.

\begin{figure}[h]
\centering
\includegraphics[scale=0.4]{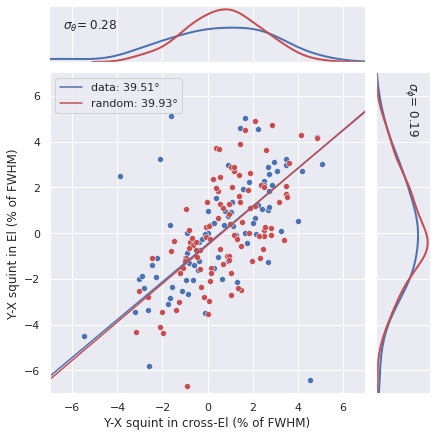}
\caption{Comparison of the measured (blue) and simulated (red) on-sky polarization beam squint. Simulations were performed assuming a random tilt of the grid with a gaussian distribution in the two orthogonal
directions. The line computed from the fit shows the direction of the beam squint in the sky.}
\label{fig:squint}       % Give a unique label
\end{figure}

Since the inaccuracy in the mounting of the grid could be one of the causes of the scatter, we tried to reproduce this effect using the GRASP model to simulate the squint. We computed the beam squint in the focal plane for a random tilt of the grid around its axes and we quantified the offset of the beam in the focal plane to make a comparison with measured data. We assumed two independent gaussian distributions for the tilt in the two perpendicular directions because they are decoupled due to the mechanical construction. 
By optimizing the values of the two independent standard deviations to match the data, we found that a gaussian distribution with a standard deviation $\sigma =$~0.28$^\circ$ in one direction and $\sigma =$~0.19$^\circ$ in the orthogonal one can give a similar scatter of the beam squint, as shown in Fig.~\ref{fig:squint}. Therefore, inaccuracies in the grid mounting could be responsible for the observed beam squint scatter.

\section{Conclusions}
\label{sec:conclusions}
The main conclusions of the study are that the SIS mixers allow an extension of the upper limit of the IF band to at least 18 GHz, the RF bandwidth can be widened to 580–735 GHz and the limitation of the current polarimetric performance can be explained by deviations in the grid mounting angle for the beam squint and by the presence of the grid in combination with the mirrors for the cross-polarization level.

% BibTeX or Biber users please use (the style is already called in the class, ensure that the "woc.bst" style is in your local directory)
% \bibliography{biblio}

\end{document}